\PassOptionsToPackage{numbers, sort&compress}{natbib}
\documentclass[unnumsec,webpdf,contemporary,large]{oup-authoring-template}%
\setcitestyle{square,comma}
\graphicspath{{Fig/}}

\theoremstyle{thmstyleone}%
\theoremstyle{thmstyletwo}%
\theoremstyle{thmstylethree}%

\usepackage{booktabs}
\usepackage[nodayofweek,level]{datetime}
\newcommand{\duallabeleddatasetdate}{\formatdate{23}{12}{2024}}
\usepackage{graphicx}
\usepackage{comment}

\begin{document}

\journaltitle{Preprint}
\DOI{}
\copyrightyear{2025}
\pubyear{2025}
\access{Preprint}
\appnotes{Paper}

\firstpage{1}

%\subtitle{Subject Section}

\title[Impact of Labeling and Image Inaccuracy on Federated, Centralized and Local Learning]{Impact of Labeling Inaccuracy and Image Noise on Tooth Segmentation in Panoramic Radiographs using Federated, Centralized and Local Learning}

%\begin{comment}
\author[1,$\ast$]{Johan Andreas Balle Rubak\ORCID{0009-0003-9229-1844}}
\author[1]{Khuram Naveed\ORCID{0000-0001-8286-6139}}
\author[1]{Sanyam Jain\ORCID{0009-0004-0699-025X}}
\author[2]{Lukas Esterle\ORCID{0000-0002-0248-1552}}
\author[3]{Alexandros Iosifidis\ORCID{0000-0003-4807-1345}} 
\author[1]{Ruben Pauwels\ORCID{0000-0002-9462-7546}}

\authormark{Johan Andreas Balle Rubak et al.}

\address[1]{\orgdiv{Department of Dentistry and Oral Health}, \orgname{Aarhus University}, \orgaddress{\street{Vennelyst Boulevard 9}, \postcode{8000}, \state{Aarhus}, \country{Denmark}}}
\address[2]{\orgdiv{Department of Electrical- and Computer Engineering}, \orgname{Aarhus University}, \orgaddress{\street{Finlandsgade 22}, \postcode{8200}, \state{Aarhus}, \country{Denmark}}}
\address[3]{\orgdiv{Faculty of Information Technology and Communications Sciences}, \orgname{Tampere University}, \orgaddress{\street{Kalevantie 4}, \postcode{33100}, \state{Tampere}, \country{Finland}}}

\corresp[$\ast$]{Corresponding author. \href{email:joru@dent.au.dk}{joru@dent.au.dk}}
%\end{comment}

%\received{Date}{0}{Year}
%\revised{Date}{0}{Year}
%\accepted{Date}{0}{Year}

%\editor{Associate Editor: Name}

%\abstract{
%\textbf{Motivation:} .\\
%\textbf{Results:} .\\
%\textbf{Availability:} .\\
%\textbf{Contact:} \href{name@email.com}{name@email.com}\\
%\textbf{Supplementary information:} Supplementary data are available at \textit{Journal Name}
%online.}

\abstract{
\textbf{Objectives:} Federated learning (FL) may mitigate privacy constraints, heterogeneous data quality, and inconsistent labeling in dental diagnostic AI. We compared FL with centralized (CL) and local learning (LL) for tooth segmentation in panoramic radiographs across multiple data corruption scenarios.\\
\textbf{Methods:} An Attention U‑Net was trained on 2066 radiographs from six institutions across four settings: baseline (unaltered data); label manipulation (dilated/missing annotations); image‐quality manipulation (additive Gaussian noise); and exclusion of a faulty client with corrupted data. FL was implemented via the Flower AI framework. Per-client training- and validation loss trajectories were monitored for anomaly detection and a set of metrics (Dice, IoU, HD, HD95 and ASSD) was evaluated on a hold‑out test set. From these metrics significance results were reported through Wilcoxon signed-rank test. CL and LL served as comparators.\\
\textbf{Results:} 
\emph{Baseline:} FL achieved a median Dice of 0.94889 (ASSD: 1.33229), slightly better than CL at 0.94706 (ASSD: 1.37074) and LL at 0.93557–0.94026 (ASSD: 1.51910-1.69777). 
\emph{Label manipulation:} FL maintained the best median Dice score at 0.94884 (ASSD: 1.46487) versus CL’s 0.94183 (ASSD: 1.75738) and LL’s 0.93003–0.94026 (ASSD: 1.51910-2.11462). 
\emph{Image noise:} FL led with Dice at 0.94853 (ASSD: 1.31088); CL scored 0.94787 (ASSD: 1.36131); LL ranged from 0.93179–0.94026 (ASSD: 1.51910-1.77350). 
\emph{Faulty‐client exclusion:} FL reached Dice at 0.94790 (ASSD: 1.33113) better than CL’s 0.94550 (ASSD: 1.39318). Loss‐curve monitoring reliably flagged the corrupted site.\\
\textbf{Conclusions:} FL matches or exceeds CL and outperforms LL across corruption scenarios while preserving privacy. Per-client loss trajectories provide an effective anomaly-detection mechanism and support FL as a practical, privacy-preserving approach for scalable clinical AI deployment.}


% \boxedtext{
% \begin{itemize}
% \item Key boxed text here.
% \item Key boxed text here.
% \item Key boxed text here.
% \end{itemize}}

\maketitle

\section{Introduction}
Artificial intelligence (AI) techniques are currently transforming and innovating diverse fields such as healthcare, finance, and logistics \cite{01, 02}. In particular in the medical domain, AI has demonstrated considerable potential in improving diagnostic accuracy and patient treatment outcomes \cite{03, 04, 05}. However, despite these benefits, especially for automating repetitive tasks, the integration of AI technology into clinical practice remains limited by challenges related to data collection and sharing \cite{challenges1, challenges2, challenges3}.

The most straightforward method for training an AI model is to use locally available data, a paradigm referred to as Local Learning (LL). Yet, to achieve an optimal and generalized model that is applicable across various hospitals, centers, and clinics (hereafter referred to as "clients"), data must ideally be aggregated in a centralized manner, which enables centralized learning (CL) \cite{cl}. This approach, however, is often restricted by regulatory, user-preference, and data volume constraints, as well as ethical and legal privacy concerns (e.g., compliance with the General Data Protection Regulation (GDPR)) \cite{gdpr}. Therefore, a novel approach is required that can balance high-performance model training with strict adherence to data privacy regulations.

Federated Learning (FL) offers such a solution by enabling multiple clients to collaboratively train a shared model while keeping the data on-site. Unlike a centralized approach, which would require both data and the associated computational load to be gathered in one location, FL distributes the training workload across each data holder, making it especially attractive for medical imaging. FL allows institutions to benefit from diverse and extensive datasets without compromising patient privacy or data security \cite{SCHNEIDER, fl3, fl4}. Nevertheless, FL introduces new challenges: each client must possess adequate computational resources, a requirement that may entail significant initial economic investment, and issues related to data heterogeneity and inconsistent labeling may further complicate model performance \cite{fl1}. Understanding these factors is essential to evaluate FL’s integration into medical applications and to identify elements that may either facilitate or hinder its clinical adoption.

The primary objective of this study is to investigate FL's potential to enhance segmentation of dental anatomical structures in panoramic radiographs, while addressing some of the specific challenges associated with its implementation. To this end, we: (1) establish an FL pipeline for simulating a central server with distinct client nodes housing local data repositories; (2) Compare the performance of FL models with those trained via CL or LL; and (3) Assess the sensitivity of FL models to the variations in image quality across clients and labeling inconsistencies in order to mimic real-world scenario.
Our hypothesis is that FL can significantly improve performance compared with LL and even rival the effectiveness of CL.

\section{Data and Methods}
\subsection{Study Design} 
Panoramic radiography provides a comprehensive two-dimensional overview of the dentition and surrounding maxillofacial structures, playing a crucial role in diagnostic assessment and treatment planning \cite{PR1}. While manual delineation of individual teeth is labor-intensive and subject to inter- and intra-observer variability, automated segmentation solutions provide a more standard and reliable alternative \cite{PR2, PR3}. This study aims to investigate the potential of FL for training neural networks to segment dental anatomical structures (i.e., teeth) in comparison to the more conventional training paradigms, namely LL and CL. To this end, we evaluate the trained models through both local validation data and centralized test data. The experiments were repeated under four conditions: (1) baseline models with unaltered data, (2) models affected by label manipulation for one client (3) models affected by image quality degradation for one client, and (4) models with one faulty client excluded. This will be discussed further in more details under the \textit{Training procedure}.

\subsection{Data}
The dataset used in this study is the open-source \textit{A dual labeled dataset} \cite{duallabeled} (the version of the dataset was downloaded the \duallabeleddatasetdate), that comprises of 2066 panoramic radiographs collected from three hospitals (China-Japan Union Hospital of Jilin University, Wuxi Stomatology Hospital, and People’s Hospital of Zhengzhou) and three dental clinics between January 1, 2015, and December 31, 2023. Imaging equipment was sourced from various manufacturers, including Orthophos XG (Dentsply Sirona, Germany), Planmeca ProMax (Planmeca, Finland), and Bondream 1020 (Bondream, China). The study population ($989$ males and $1011$ females) consisted of individuals aged 16 years and older, with a mean age of $36.6$ years. Data annotation was performed using Labelme (v5.4.1) by four dentists concurrently; who delineated the teeth using polygonal annotations with zooming enabled for accuracy. Notably, dental implants were not annotated. After delineation, teeth were assigned numbering and status labels according to the Federation Dentaire Internationale (FDI) notation, ranging from $11$ to $48$, with $91$ representing supernumerary teeth, resulting in $33$ distinct numbering labels. Additionally, seven status labels were applied \cite{duallabeled}; these are not used within the current study.

\subsection{Model}
We employ a U-Net with integrated attention blocks (Attention U-Net) \cite{attenunet} across all learning paradigms to ensure a fair evaluation. The attention Unet was used because of its high performance in dental radiograph segmentation. As opposed to the conventional U-Net \cite{unet} that tends to process all image regions indiscriminately, the Attention U-Net incorporates attention gates to allow the model to focus on the most relevant target structures while suppressing irrelevant regions. The network architecture consists of five encoding blocks and five decoding blocks, with attention blocks integrated at the skip connections between them. Each attention gate dynamically weights the feature maps to emphasize critical anatomical details and suppress irrelevant information, via a soft attention mechanism. In brief, the concatenated feature maps are first processed by a $1\times1$ convolution that reduces the number of channels by a factor of eight, followed by a sigmoid activation to generate preliminary attention coefficients. A subsequent $1\times1$ convolution then restores the original channel dimensions, and a further sigmoid activation produces the final attention weights. These weights, which range from 0 to 1, are multiplied element-wise with the input feature map, thereby modulating the influence of each spatial region according to its relevance.

\begin{figure}[htbp]
  \centering
  \includegraphics[width=0.45\textwidth]{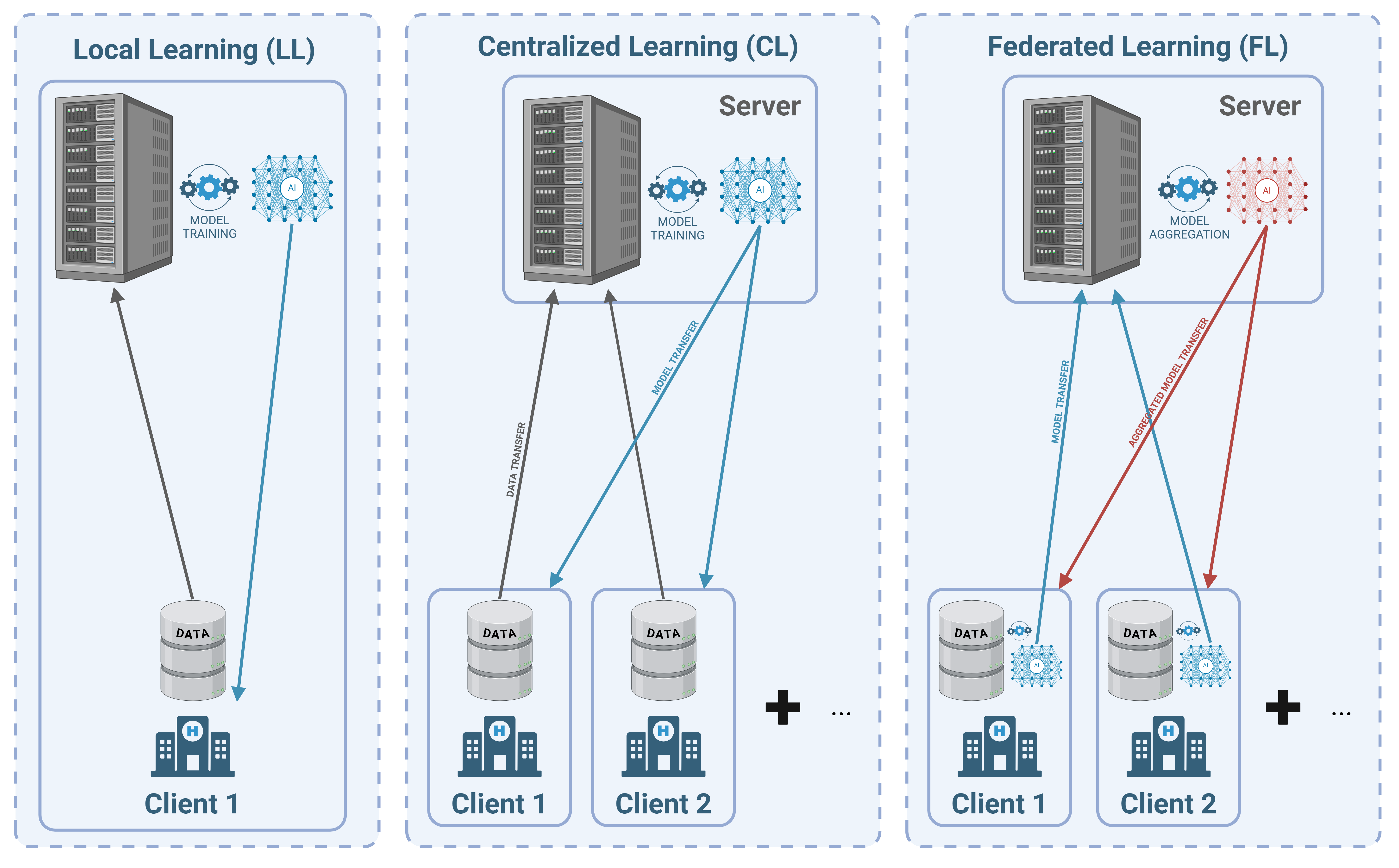}
  \caption{Learning Paradigms, to the left Local Learning, in the middle Centralized Learning and to the right Federated Learning.}
  \label{fig:learningparadigms}
\end{figure}

\subsection{Learning Paradigms} 
This study investigates three learning paradigms (illustrated in Figure \ref{fig:learningparadigms}).
\begin{itemize}
    \item \textbf{Local Learning (LL):} Training is performed solely on a single client’s local data.
    \item \textbf{Centralized Learning (CL):} Data from all participating clients is pooled at a central location for training, typically enhancing model's generalizability. 
    \item \textbf{Federated Learning (FL):}  Clients do not exchange sensitive data directly but instead share model parameters, enabling joint training on diverse multi-center and multi-vendor datasets while preserving data privacy.
\end{itemize}
The proposed FL process comprises the following steps: 
\begin{enumerate} 
    \item Initialize a global model on the server with random parameters. 
    \item Distribute the global model to the connected clients. \item Each client trains the model locally for a predetermined number of epochs (without necessarily achieving full convergence). 
    \item Clients return their locally trained models to the server. 
    \item The server aggregates these models often using federated averaging (FedAvg) \cite{fl3, fl5} that computes a weighted average based on the number of training examples per client. \item Steps 2 through 5 are repeated for several rounds until convergence, with the updated global model redistributed to all clients at the start of each round.
\end{enumerate}

To implement this process, we employed Flower AI \cite{flower}, an open-source framework for building federated learning pipelines.

\subsection{Experimental configurations}
\begin{figure}[ht]
  \centering

  \begin{minipage}{0.45\textwidth}
    \centering
    \includegraphics[width=\linewidth]{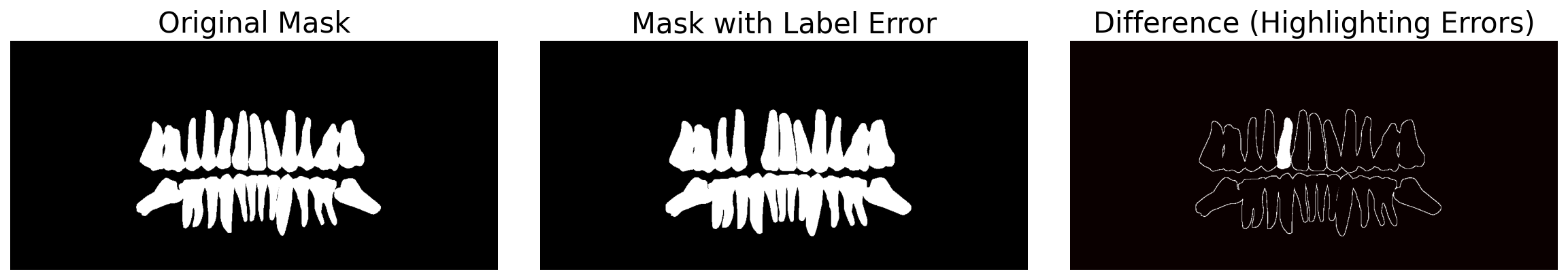}\\[0.5ex]
    \textbf{(a)}
    \label{fig:label_errors}
  \end{minipage}%
  \hspace{0.5cm}%
  \begin{minipage}{0.45\textwidth}
    \centering
    \includegraphics[width=\linewidth]{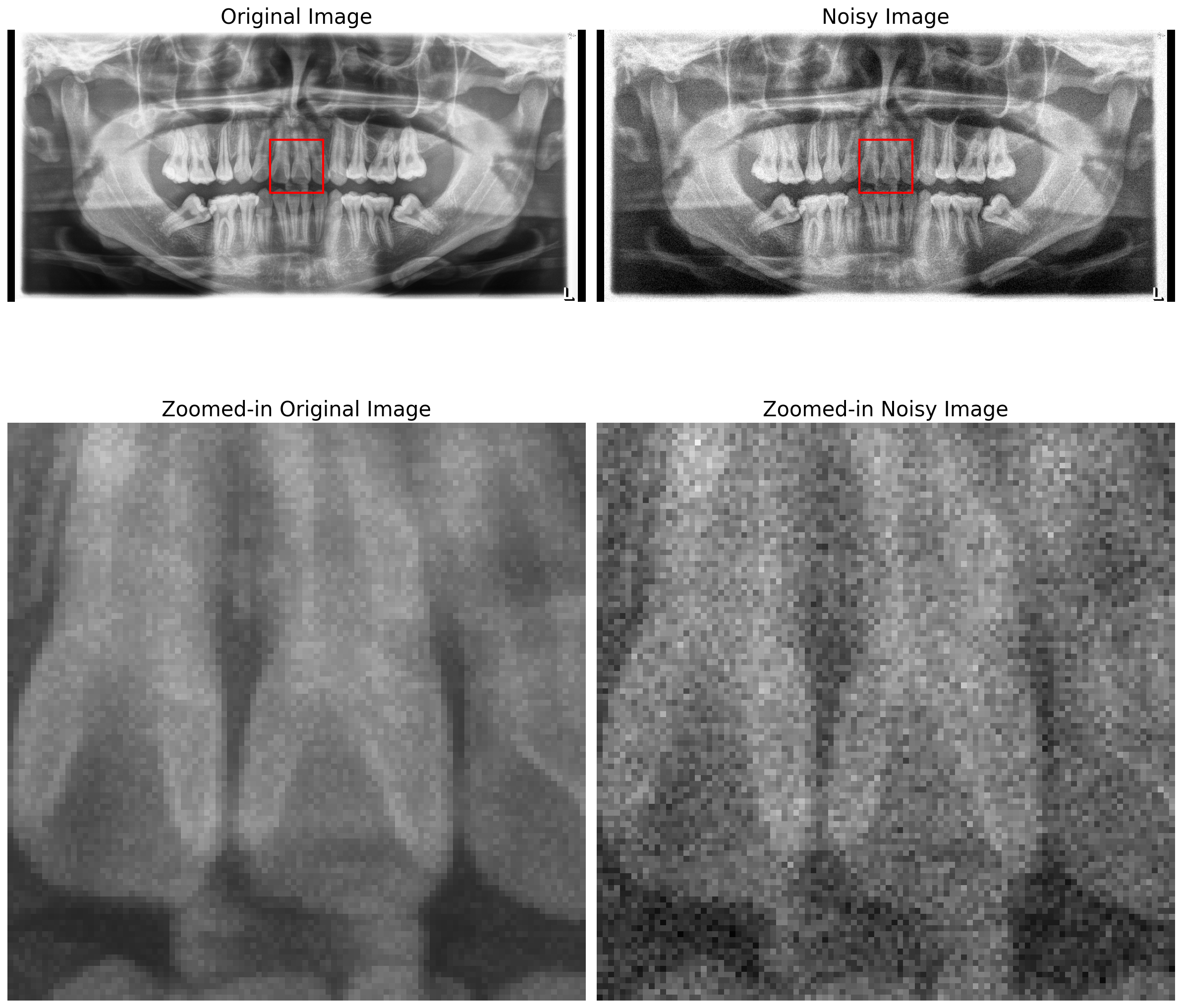}\\[0.5ex]
    \textbf{(b)}
    \label{fig:image_errors}
  \end{minipage}

  \caption{%
    Comparison between label errors and image errors. (a) Label errors in the mask, showing dilation of the label and a missing tooth. (b) Image quality degradation to one of the clients through addition of noise. Red boxes at top images act as bounding box for the bottom images.%
  }
  \label{fig:errors_comparison}
\end{figure}

All learning paradigms were carried out four times in different experimental configurations with the intention of investigating the performance outcome. Faulty clients were introduced as the ones whose data integrity is compromised, either through incorrect annotations or degraded image quality. Label and image quality manipulations were grounded in plausible issues that could potentially arise from a single client \cite{labelerror1, labelerror2, imagequality1, imagequality2, imagequality3}. However, these manipulations were intentionally exaggerated to elicit a discernible and statistically significant response. \textbf{(1) Baseline Models:} Data remained un-altered. \textbf{(2) Label Manipulation:} To mimic poorly performed, rapid annotations, where annotators roughly trace tooth outlines rather than follow precise edges, we applied mask dilation to the ground-truth labels for one client, an example of this can be found in Figure \ref{fig:errors_comparison}. Dilation used binary kernels of size $3$, $5$, $7$, and 11 randomly per radiograph to exaggerate boundary inaccuracy. Additionally, each image had a $10\%$ probability of omitting one randomly selected tooth mask, simulating occasional missed annotations. \textbf{(3) Image Quality Manipulation:} For one client, image quality was degraded by adding Gaussian noise with a mean of $\mu=0$ and a standard deviation of $\sigma=25$ (considering an 8-bit grey scale). A configuration involving five clients was used throughout these experiments (see Figure \ref{fig:errors_comparison} for error induced examples). \textbf{(4) Faulty Client Exclusion:} For this configuration the faulty client was excluded from participating in training the CL and FL model.

\subsection{Training procedure}
At first, the dataset was split with $10\%$ (206 images and labels) randomly designated as the test set, used consistently across all paradigms and experimental configurations. Note that LL$_1$ through LL$_4$ were each trained only once, since their input data remained unchanged across configurations, whereas LL$_0$ was retrained for every configuration (except Faulty Client Exclusion). Data management proceeded as follows (illustrated in Figure \ref{fig:trainingprocedure}):

\begin{figure*}[htbp]
  \centering
  \includegraphics[width=\textwidth]{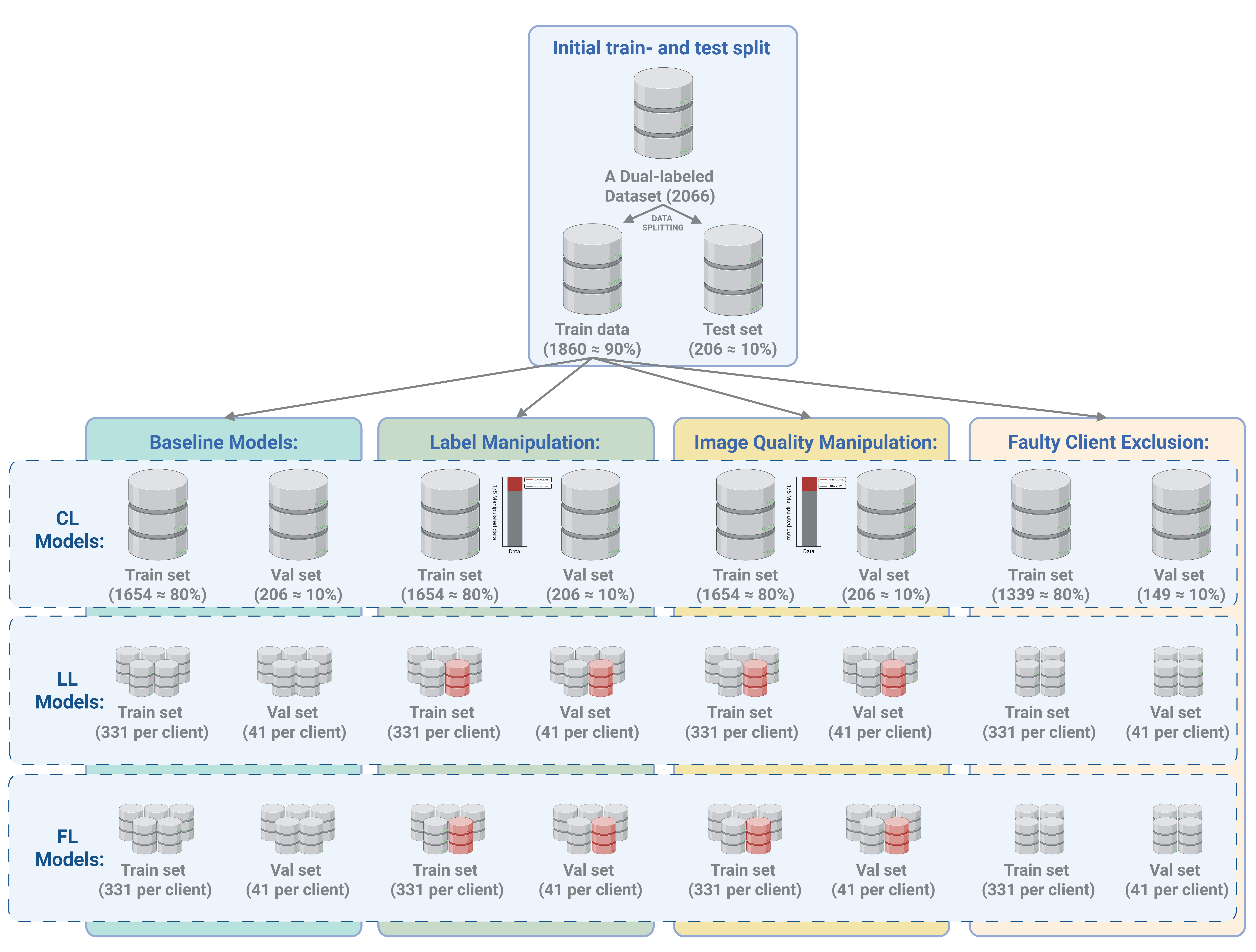}
  \caption{Data management, showing initial split of test data, following partitioning into training (train) and validation (val) sets for each learning paradigm and experimental configuration. Grey depicts untouched data, while red shows manipulated data or labels.}
  \label{fig:trainingprocedure}
\end{figure*}

\begin{enumerate}
    \item \textbf{Baseline models}
    \begin{enumerate} 
        \item CL Models: The remaining data was partitioned into approximately $10\%$ validation data (206 images and labels) and $80\%$ training data (1654 images and labels).
        \item LL Models: A random sample comprising $\frac{1}{5}$ of the remaining data was made for each client and then split into approximately $10\%$ validation data ($41$ images and labels) and 80\% training data ($331$ images and labels). 
        \item FL Models: An IidPartitioner was used to randomly divide (same random seed as for LL) the remaining data into five partitions, with each partition subsequently split into approximately $10\%$ validation data ($41$ images and labels) and $80\%$ training data ($331$ images and labels).
    \end{enumerate}
    \item \textbf{Label manipulation}
    \begin{enumerate}
        \item CL Models: Label manipulation was applied to $\frac{1}{5}$ of the remaining data, equivalent to the partition belonging to client $0$, prior to splitting into validation and training sets.
        \item LL Models: Label manipulation was applied to the data belonging to client $0$.
        \item FL Models: After partitioning the data into five, label manipulation was applied exclusively to the data of client $0$.
    \end{enumerate}
    \item \textbf{Image manipulation}
    \begin{enumerate}
        \item CL Models: Image manipulation was applied to $\frac{1}{5}$ of the remaining data, equivalent to the partition belonging to client 0, before splitting into validation and training sets. 
        \item LL Models: Image manipulation was applied to the data belonging to client 0. 
        \item FL Models: After partitioning the data into five, image manipulation was applied solely to the data of client 0.
    \end{enumerate}
    \item \textbf{Faulty client exclusion}
    \begin{enumerate}
        \item CL Models: After excluding the data belonging to client 0, the remaining data was partitioned into approximately $10\%$ validation data (149 images and labels) and $90\%$ training data ($1339$ images and labels).  
        \item LL Models: Same as the ones trained for baseline.
        \item FL Models: Only clients $1$-$4$ were allowed to participate.
    \end{enumerate}
\end{enumerate}

The most common image dimensions in the dataset were $2800\times 1316$ pixels (aspect ratio $\approx$ 2.125). Prior to training, images were resized to a height of $512$ pixels, with the width adjusted to $1088$ pixels to preserve the aspect ratio. Images with a width smaller than 1088 pixels were zero-padded, whereas those with a larger width were cropped. Training was performed on an NVIDIA RTX $6000$ Ada Generation GPU with 48 GB of graphics memory, which allowed a maximum batch size of $4$ in the federated setup. The training process employed the AdamW optimizer with a learning rate of $1\times10^{-4}$ and utilized a Dice loss function. All models were trained for $50$ epochs in total; for FL models, these epochs were distributed over $5$ rounds ($10$ epochs per round). The Flower framework was used to simulate the federated setup, where all clients participated in the aggregation process based on FedAvg after each round, except for the Faulty Client Exclusion configuration. Tracking the training process was done using validation data after each epoch while logging metrics.

\subsection{Performance metrics and statistical analysis}
Model performance was evaluated primarily using Dice score (Sørensen–Dice coefficient) computed on both local validation data and the centralized test set.  
The Dice score quantifies the similarity between the ground truth mask \(G\) and the predicted mask \(P\) by measuring their overlap:
\begin{equation}
    \mathrm{Dice}(G,P) \;=\; \frac{2\,|G \cap P|}{\,|G| + |P|\,},
\end{equation}
where \(|\cdot|\) denotes the number of positive pixels in a mask. The Dice score ranges from 0 (no overlap) to $1$ (perfect overlap).
The Dice \emph{loss} used during training is defined as
\begin{equation}
    \mathcal{L}_{\mathrm{Dice}} \;=\; 1 - \mathrm{Dice}(G,P),
\end{equation}
and should not be confused with the Dice \emph{score} above.

To give an extended understanding of the model performance, additional metrics were computed, including the Intersection over Union (IoU), also known as the Jaccard index. IoU quantifies the ratio of the overlapping area to the union of the ground truth ($G$) and predicted ($P$) segmentations: 
\begin{equation}
\mathrm{IoU}(G,P)=\frac{|G \cap P|}{|G \cup P|}
\end{equation}

As a complementary boundary‐focused metric, the Hausdorff distance (HD) was calculated on the centralized test set. For two binary masks, the Hausdorff distance is defined as:
\begin{equation}
\mathrm{HD}(G, P)
= \max\Bigl\{
    \max_{g\in G}\,\min_{p\in P}d(g,p)\;,\;
    \max_{p\in P}\,\min_{g\in G}d(p,g)
\Bigr\}
\end{equation}

where $d(g,p)$ represents the Euclidean distance between point $g$ in the ground truth set $G$ and point $p$ in the predicted segmentation set $P$.  
The $95$th percentile Hausdorff distance (HD$95$) was also reported, which is less sensitive to outlier errors by excluding the largest 5\% of point-to-point distances.  

Finally, the Average Symmetric Surface Distance (ASSD) measures the mean bidirectional distance between the predicted and ground truth surfaces:
\begin{equation}
\resizebox{0.9\columnwidth}{!}{%
  $\displaystyle
  \mathrm{ASSD}(G,P) = \frac{\sum_{g\in G_s} \min_{p\in P_s} d(g,p) \;+\;
  \sum_{p\in P_s} \min_{g\in G_s} d(p,g)}{|G_s|+|P_s|}$
}
\end{equation}
where $G_s$ and $P_s$ denote the sets of surface points in $G$ and $P$, respectively. For both HD, HD95 and ASSD the distance unit is pixels, since a physical scaling factor is not applicable for this type of image.

To assess whether the metric distributions met the assumption of normality, Shapiro–Wilk tests were applied to data from each learning paradigm under every experimental configuration. As none of the groups passed the normality criterion ($p<0.05$ for all), the analysis were conducted with non‐parametric tests.

Descriptive statistics are reported as median ($50$th percentile), interquartile range (IQR; $25$th–$75$th percentiles), and $95$th percentile values for all metrics. To evaluate statistical significance of pairwise differences, the Wilcoxon signed‐rank test (two‐sided) was employed. Specifically, the following comparisons were performed:

\begin{itemize}
  \item Within‐configuration comparisons between paradigms: LL$_0$ vs. CL, LL$_0$ vs. FL, and CL vs. FL, for each of the four configurations.
  \item Within‐paradigm comparisons across configurations: baseline vs. label manipulation, baseline vs. image quality manipulation, baseline vs. faulty client exclusion, and all other pairwise configuration combinations, for each paradigm.
\end{itemize}

To control the family‐wise error rate due to multiple testing, a Bonferroni correction was applied to the significance threshold $\alpha$. Adjusted \( \alpha_{\text{corr}} = 0.05 /N_{\text{tests}} \) was used, where \( N_{\text{tests}} \) is the total number of comparisons performed. In total, 18 tests were performed; therefore, reported \(p\)-values falling below the corrected threshold at $\alpha_{\text{corr}}=0.00278$ were considered statistically significant.

\section{Results}
\subsection{Baseline models}
The training progress for each experimental configuration and its respective learning paradigms were tracked. Training progress of the baseline models can be found in Figure \ref{fig:all_progress}.

Metrics based on test data for all models and learning paradigms are shown in Table \ref{tab:final_results} and Figure \ref{fig:all_boxplot}. FL reached the best median score for all metrics except for HD, where CL reached a better score. LL models had the poorest performance for all metrics.

\subsection{Label Manipulation}
Training progress of models with induced label errors can be found in Figure \ref{fig:all_progress}. Note that LL models are not trained again except LL$_0$, which had its data manipulated.

For each learning paradigm with induced label manipulation, FL achieved the best performance on the test data for all metrics except HD, where LL$_4$ scored best. CL was the second-best model, with LL models (especially LL$_0$ i.e. the model trained on manipulated data) generally performing the worst. Boxplots of each metric can be found in Figure \ref{fig:all_boxplot}.

\subsection{Image Quality Manipulation}
Training progress of models with induced image quality errors can be found in Figure \ref{fig:all_progress}. Note that LL models were not trained again except LL$_0$ which had its data manipulated.

Among the learning paradigms with induced image quality manipulation, FL exhibited the best performance on the test data for all metrics except HD, where CL reached a better score. For metrics other than HD, CL showed the second-best performance and LL models the worst, with LL$_0$ performing poorest. Boxplot of metrics can be found in Figure \ref{fig:all_boxplot}.

\subsection{Faulty Client Exclusion}
\begin{figure*}[htbp] % dedicated float page (full-page)
  \centering

  % --- top-left
  \begin{minipage}[b]{0.5\textwidth}
    \centering
    \includegraphics[width=\linewidth,height=0.46\textheight,keepaspectratio]{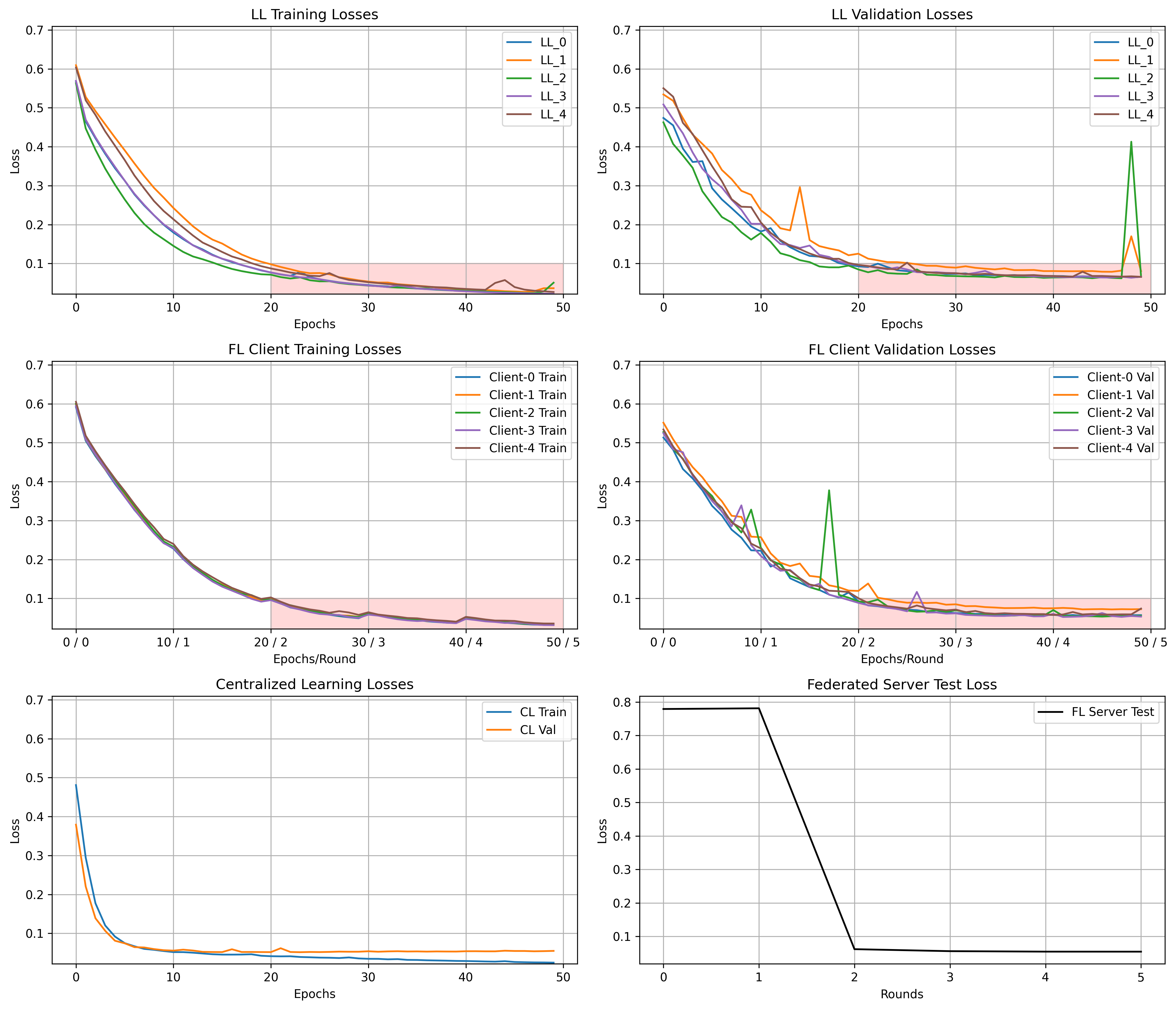}
    \\[4pt]\footnotesize (a) Baseline models (LL, FL, CL).
  \end{minipage}\hfill
  % --- top-right
  \begin{minipage}[b]{0.5\textwidth}
    \centering
    \includegraphics[width=\linewidth,height=0.46\textheight,keepaspectratio]{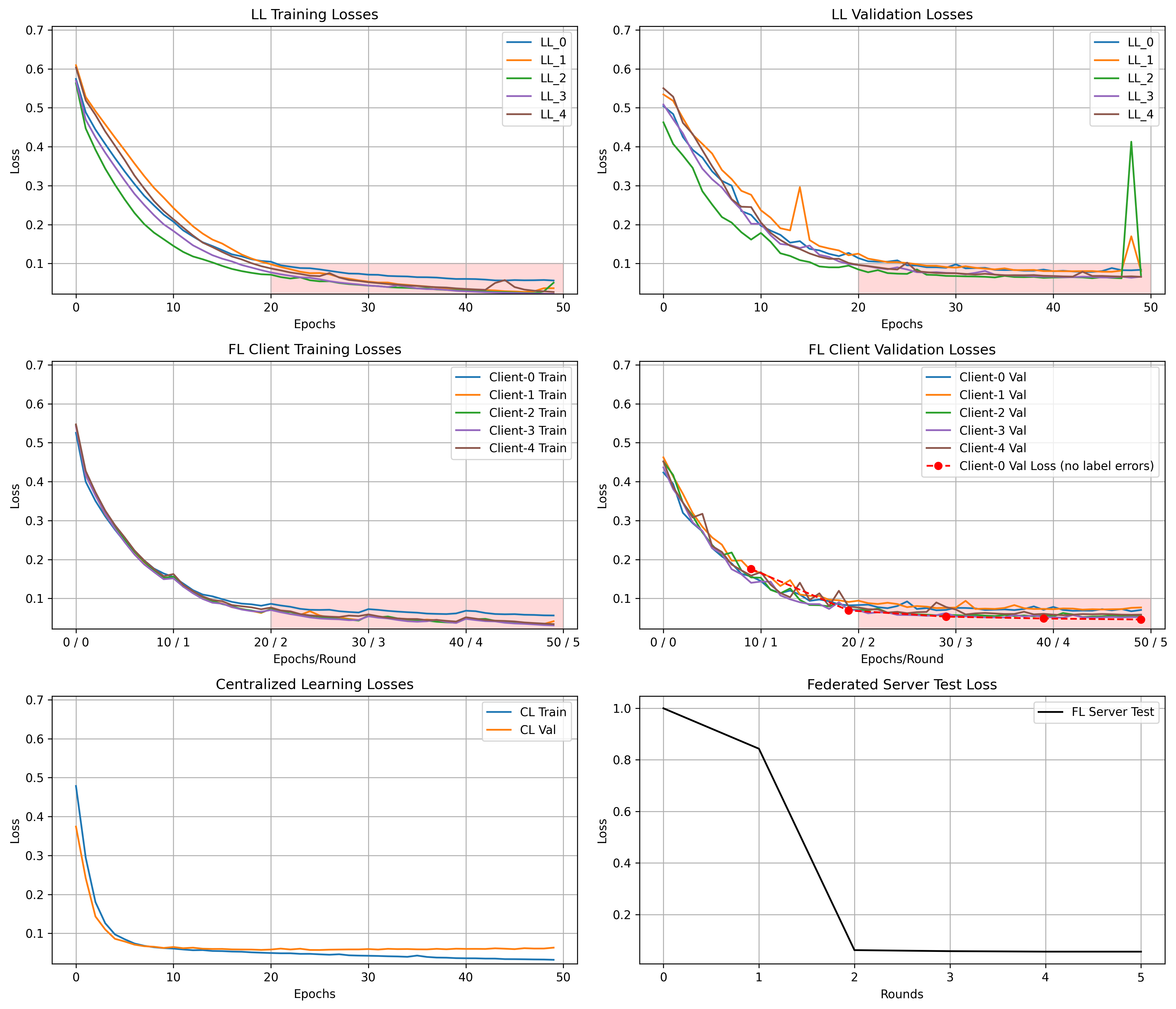}
    \\[4pt]\footnotesize (b) Label manipulation (LL, FL, CL).
  \end{minipage}

  \vspace{8pt}

  % --- bottom-left
  \begin{minipage}[b]{0.5\textwidth}
    \centering
    \includegraphics[width=\linewidth,height=0.46\textheight,keepaspectratio]{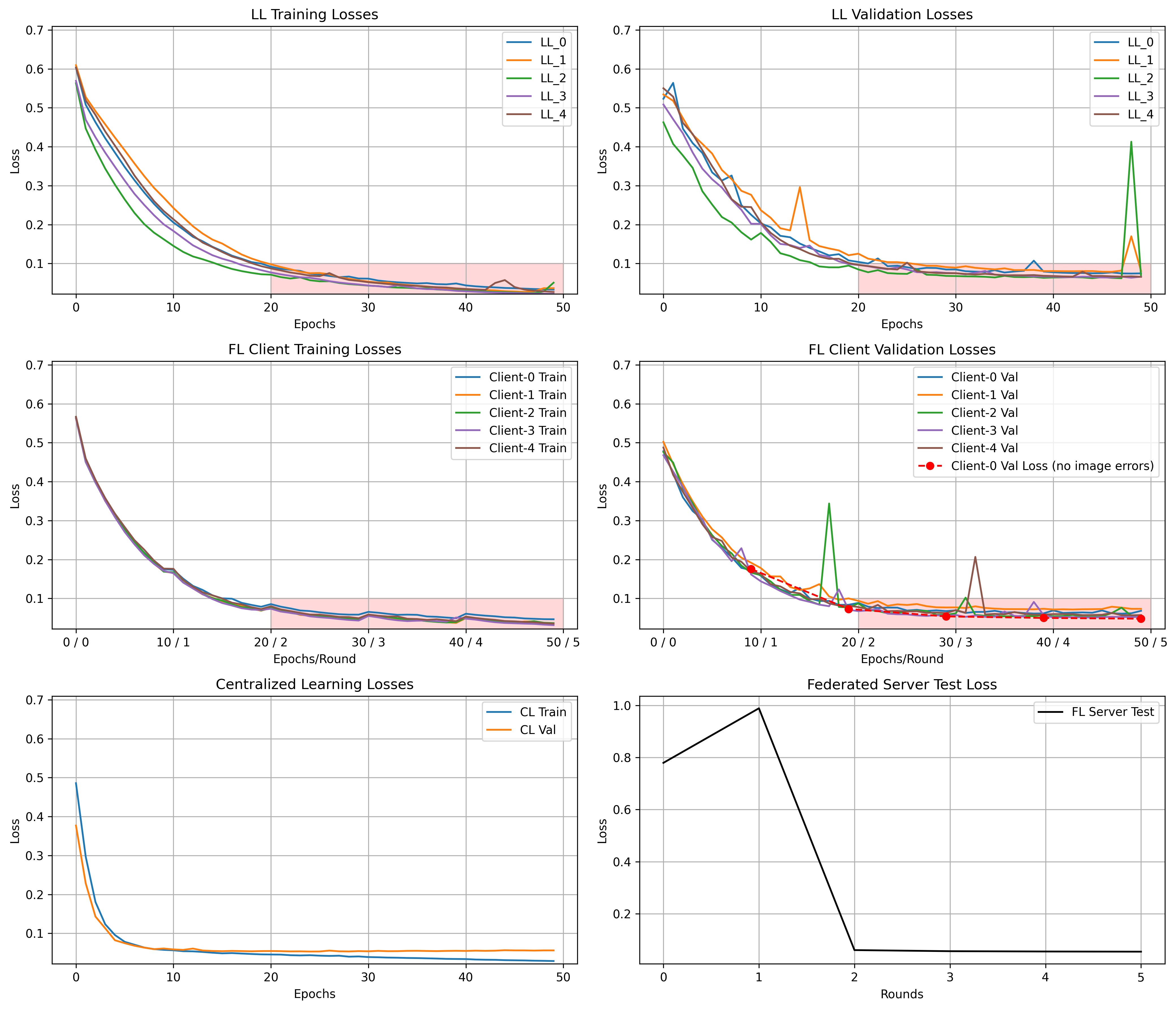}
    \\[4pt]\footnotesize (c) Image-quality manipulation (LL, FL, CL).
  \end{minipage}\hfill
  % --- bottom-right
  \begin{minipage}[b]{0.5\textwidth}
    \centering
    \includegraphics[width=\linewidth,height=0.46\textheight,keepaspectratio]{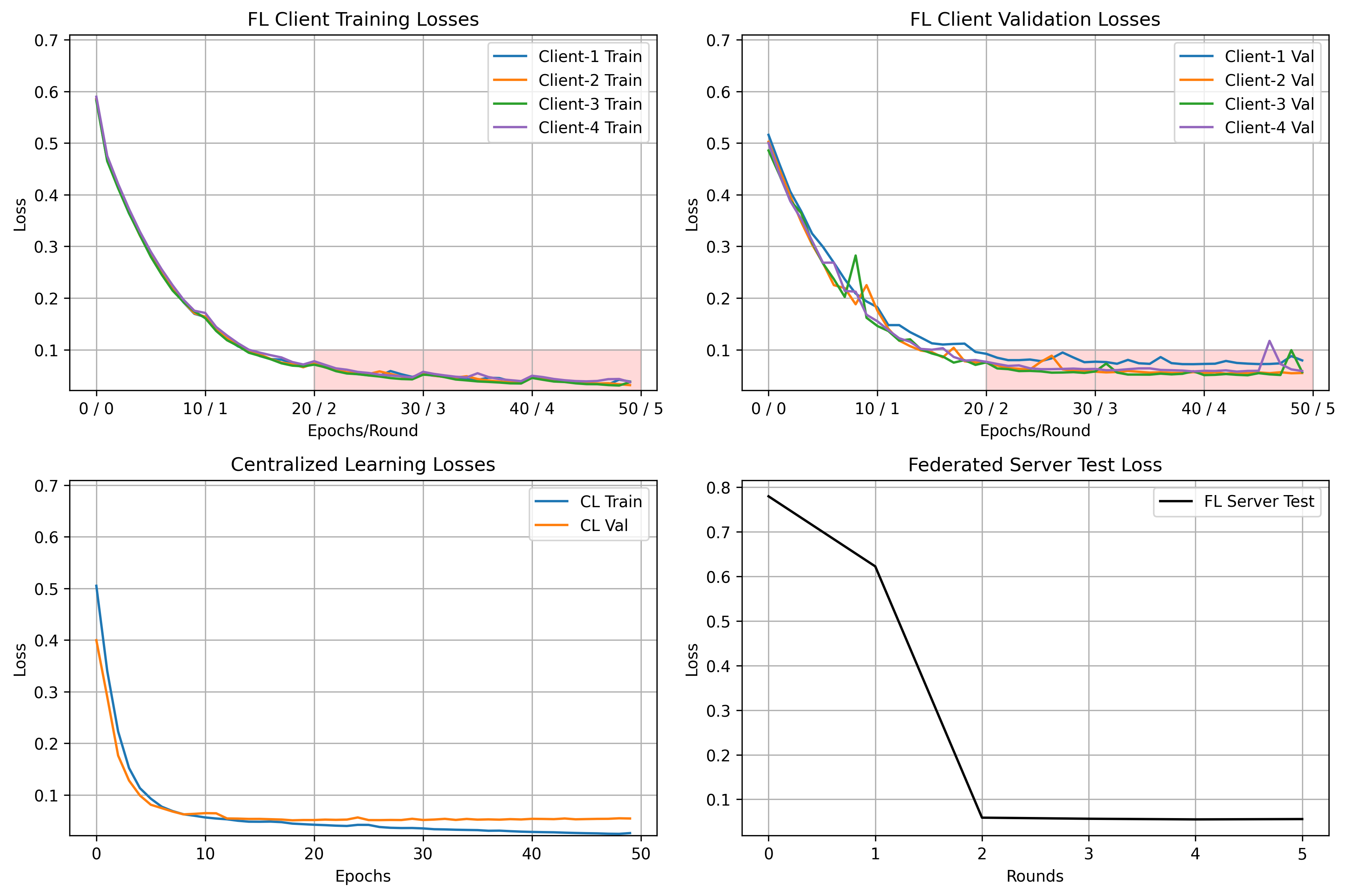}
    \\[4pt]\footnotesize (d) Faulty-client exclusion (FL and CL, client 0 excluded).
  \end{minipage}

  \caption{Training progress for all experiments. (a) Baseline; (b) Label manipulation; (c) Image-quality manipulation; (d) Faulty-client exclusion. Red boxes indicate areas to check for convergence/elevated loss, especially for LL$_0$ and client-0.}
  \label{fig:all_progress}
\end{figure*}

Training progress of CL and FL models excluding client 0 can be found in Figure \ref{fig:all_progress}; note that LL models were not trained again, since the non-excluded LL clients were already trained for the baseline configuration.

\begin{figure*}[htbp]
  \centering
  \includegraphics[width=0.9\textwidth]{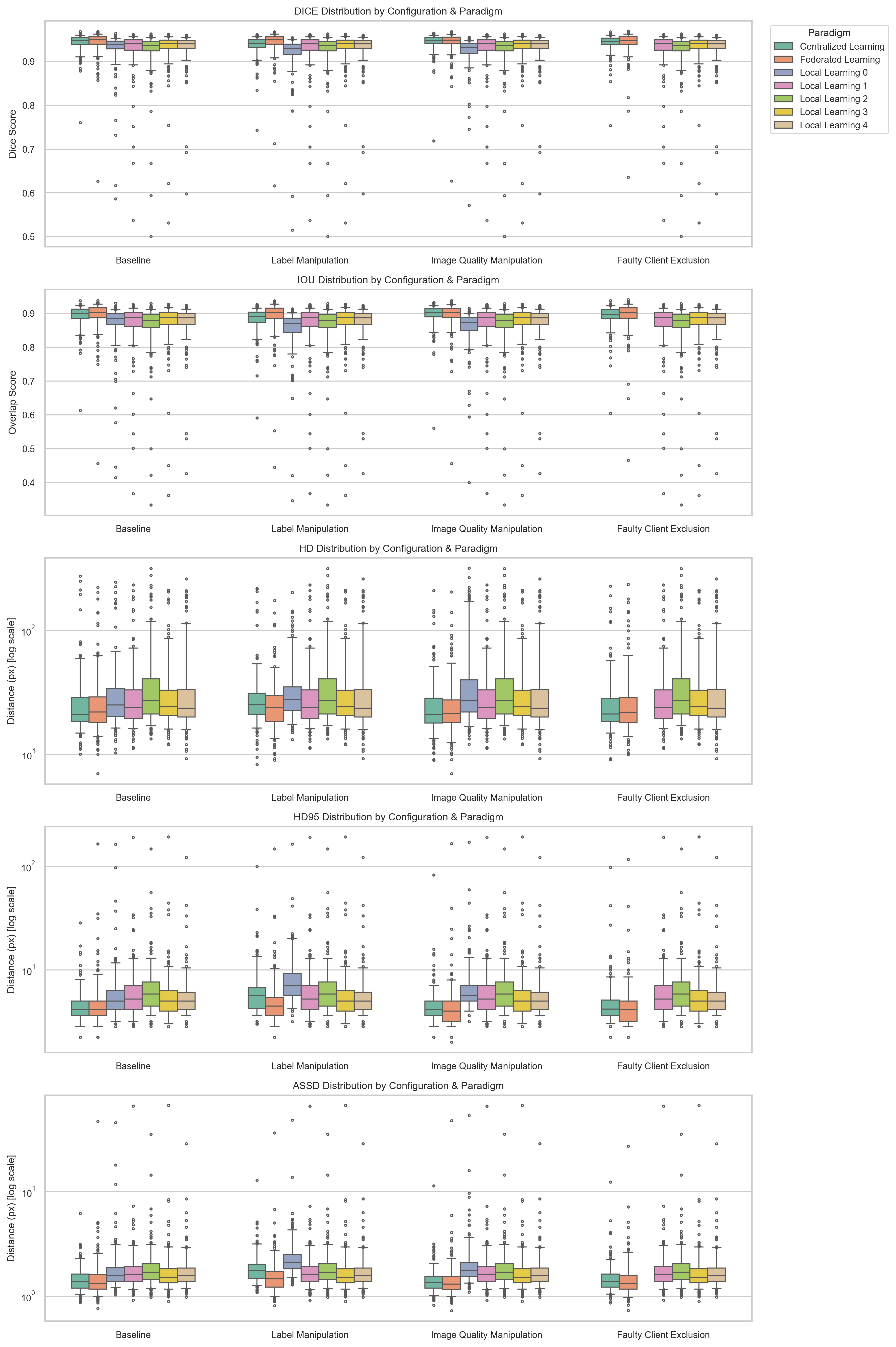}
  \caption{Grouped boxplots with whiskers representing the [5, 95] percentiles of all metrics across centralized, federated and local learning. Outliers not within these percentiles are represented as dots.}
  \label{fig:all_boxplot}
\end{figure*}

After faulty client exclusion, FL showed the best performance for all metrics except HD, where CL reached a better score. For metrics other than HD, CL showed the second-best performance, whereas LL models performed the worst. Boxplots of metrics can be found in Figure \ref{fig:all_boxplot}.

\subsection{Statistical Comparison of Results}
Statistical analysis using the two-sided Wilcoxon signed-rank test was performed for both within-configuration and within-paradigm comparisons. A comprehensive presentation of the significance test results can be found in Table \ref{tab:within-configuration} and \ref{tab:within-paradigm} in the Appendix. Significant differences were observed across almost all configurations, with the exception of comparisons between Federated Learning (FL) and Centralized Learning (CL) based on Hausdorff Distance (HD), 95th percentile Hausdorff Distance (HD95), and Average Symmetric Surface Distance (ASSD) within the Baseline configuration; HD and HD95 within the Image Quality Manipulation configuration; and HD within the Faulty Client Exclusion configuration.

Within-paradigm comparisons revealed significant differences for all metrics when contrasting Baseline models with Label Manipulation models, except for HD in the FL model. No significant differences were detected between the Baseline and Image Quality Manipulation configurations, apart from the LL$_0$ model. Similarly, comparisons between Baseline and Faulty Client Exclusion models showed no statistically significant differences, although both CL and FL models demonstrated an improvement in performance when the faulty client was excluded.

\begin{table*}[htbp]
\caption{Descriptive statistics (Median, IQR, 95th percentile) for all metrics by Configuration and Paradigm on Test-set. The best median metric value for each configuration and metric is underlined.}
\label{tab:final_results}
\resizebox{\textwidth}{!}{%
\begin{tabular}{ll rrr rrr rrr rrr rrr}
\toprule
\multirow{2}{*}{Configuration} & \multirow{2}{*}{Paradigm} & \multicolumn{3}{c}{DICE} & \multicolumn{3}{c}{IOU} & \multicolumn{3}{c}{HD [px]} & \multicolumn{3}{c}{HD95 [px]} & \multicolumn{3}{c}{ASSD [px]} \\
 &  & Median & IQR & P95 & Median & IQR & P95 & Median & IQR & P95 & Median & IQR & P95 & Median & IQR & P95 \\
\midrule
\multirow{7}{*}{Baseline} & CL & 0.94706 & 0.01515 & 0.95934 & 0.89944 & 0.02730 & 0.92185 & \underline{21.09502} & 10.20708 & 60.99760 & \underline{4.12311} & 1.39445 & 8.72377 & 1.37074 & 0.43160 & 2.32418 \\
 & FL & \underline{0.94889} & 0.01645 & 0.96232 & \underline{0.90275} & 0.02972 & 0.92737 & 22.00000 & 10.93768 & 62.36985 & \underline{4.12311} & 1.39445 & 9.60976 & \underline{1.33229} & 0.44289 & 2.62177 \\
 & LL$_0$ & 0.93894 & 0.01782 & 0.95277 & 0.88491 & 0.03156 & 0.90980 & 25.01000 & 13.78110 & 71.53083 & 5.00000 & 2.20145 & 11.97776 & 1.56560 & 0.48196 & 3.22943 \\
 & LL$_1$ & 0.94010 & 0.02327 & 0.95601 & 0.88697 & 0.04123 & 0.91572 & 23.89528 & 13.50307 & 73.99574 & 5.23494 & 2.87689 & 13.02880 & 1.62033 & 0.53646 & 3.13048 \\
 & LL$_2$ & 0.93557 & 0.02166 & 0.95377 & 0.87895 & 0.03818 & 0.91162 & 27.04624 & 19.46147 & 122.04397 & 5.83095 & 3.14364 & 13.98837 & 1.69777 & 0.60567 & 3.20559 \\
 & LL$_3$ & 0.94026 & 0.01981 & 0.95626 & 0.88725 & 0.03517 & 0.91618 & 24.19711 & 12.35330 & 87.79542 & 5.00000 & 2.32456 & 10.97660 & 1.51910 & 0.48987 & 2.96390 \\
 & LL$_4$ & 0.93966 & 0.01808 & 0.95435 & 0.88619 & 0.03207 & 0.91268 & 23.53720 & 13.27045 & 114.15153 & 5.00000 & 1.95966 & 10.68782 & 1.57904 & 0.47235 & 2.92375 \\
\midrule
\multirow{7}{*}{Label Manipulation} & CL & 0.94183 & 0.01717 & 0.95577 & 0.89005 & 0.03060 & 0.91529 & 25.12961 & 10.19018 & 57.71995 & 5.65685 & 2.46556 & 13.87029 & 1.75738 & 0.53663 & 3.17018 \\
 & FL & \underline{0.94884} & 0.01686 & 0.96215 & \underline{0.90266} & 0.03044 & 0.92705 & 23.92686 & 11.44332 & 50.73865 & \underline{4.47214} & 1.77961 & 10.65346 & \underline{1.46487} & 0.51296 & 2.83928 \\
 & LL$_0$ & 0.93003 & 0.02381 & 0.94898 & 0.86920 & 0.04140 & 0.90292 & 27.52264 & 12.43628 & 90.05001 & 7.00000 & 3.54963 & 20.01874 & 2.11462 & 0.68016 & 4.48904 \\
 & LL$_1$ & 0.94010 & 0.02327 & 0.95601 & 0.88697 & 0.04123 & 0.91572 & 23.89528 & 13.50307 & 73.99574 & 5.23494 & 2.87689 & 13.02880 & 1.62033 & 0.53646 & 3.13048 \\
 & LL$_2$ & 0.93557 & 0.02166 & 0.95377 & 0.87895 & 0.03818 & 0.91162 & 27.04624 & 19.46147 & 122.04397 & 5.83095 & 3.14364 & 13.98837 & 1.69777 & 0.60567 & 3.20559 \\
 & LL$_3$ & 0.94026 & 0.01981 & 0.95626 & 0.88725 & 0.03517 & 0.91618 & 24.19711 & 12.35330 & 87.79542 & 5.00000 & 2.32456 & 10.97660 & 1.51910 & 0.48987 & 2.96390 \\
 & LL$_4$ & 0.93966 & 0.01808 & 0.95435 & 0.88619 & 0.03207 & 0.91268 & \underline{23.53720} & 13.27045 & 114.15153 & 5.00000 & 1.95966 & 10.68782 & 1.57904 & 0.47235 & 2.92375 \\
\midrule
\multirow{7}{*}{Image Quality Manipulation} & CL & 0.94787 & 0.01317 & 0.95931 & 0.90091 & 0.02380 & 0.92181 & \underline{20.94031} & 10.48371 & 56.05949 & 4.12311 & 1.39445 & 7.47960 & 1.36131 & 0.35754 & 2.22228 \\
 & FL & \underline{0.94853} & 0.01515 & 0.96145 & \underline{0.90210} & 0.02736 & 0.92576 & 21.38925 & 9.47223 & 60.28257 & \underline{4.00000} & 1.83772 & 8.04669 & \underline{1.31088} & 0.39002 & 2.42075 \\
 & LL$_0$ & 0.93179 & 0.02218 & 0.94702 & 0.87229 & 0.03868 & 0.89937 & 27.04624 & 17.87757 & 175.25294 & 5.65685 & 2.07107 & 14.19981 & 1.77350 & 0.56850 & 3.81118 \\
 & LL$_1$ & 0.94010 & 0.02327 & 0.95601 & 0.88697 & 0.04123 & 0.91572 & 23.89528 & 13.50307 & 73.99574 & 5.23494 & 2.87689 & 13.02880 & 1.62033 & 0.53646 & 3.13048 \\
 & LL$_2$ & 0.93557 & 0.02166 & 0.95377 & 0.87895 & 0.03818 & 0.91162 & 27.04624 & 19.46147 & 122.04397 & 5.83095 & 3.14364 & 13.98837 & 1.69777 & 0.60567 & 3.20559 \\
 & LL$_3$ & 0.94026 & 0.01981 & 0.95626 & 0.88725 & 0.03517 & 0.91618 & 24.19711 & 12.35330 & 87.79542 & 5.00000 & 2.32456 & 10.97660 & 1.51910 & 0.48987 & 2.96390 \\
 & LL$_4$ & 0.93966 & 0.01808 & 0.95435 & 0.88619 & 0.03207 & 0.91268 & 23.53720 & 13.27045 & 114.15153 & 5.00000 & 1.95966 & 10.68782 & 1.57904 & 0.47235 & 2.92375 \\
\midrule
\multirow{6}{*}{Faulty Client Exclusion} & CL & 0.94550 & 0.01494 & 0.95936 & 0.89663 & 0.02688 & 0.92189 & \underline{21.14232} & 9.73579 & 60.71267 & 4.18287 & 1.49347 & 8.57462 & 1.39318 & 0.41393 & 2.26050 \\
 & FL & \underline{0.94790} & 0.01694 & 0.96244 & \underline{0.90096} & 0.03061 & 0.92761 & 21.84033 & 10.61997 & 69.72217 & \underline{4.12311} & 1.83772 & 9.05066 & \underline{1.33113} & 0.41297 & 2.71956 \\
 & LL$_1$ & 0.94010 & 0.02327 & 0.95601 & 0.88697 & 0.04123 & 0.91572 & 23.89528 & 13.50307 & 73.99574 & 5.23494 & 2.87689 & 13.02880 & 1.62033 & 0.53646 & 3.13048 \\
 & LL$_2$ & 0.93557 & 0.02166 & 0.95377 & 0.87895 & 0.03818 & 0.91162 & 27.04624 & 19.46147 & 122.04397 & 5.83095 & 3.14364 & 13.98837 & 1.69777 & 0.60567 & 3.20559 \\
 & LL$_3$ & 0.94026 & 0.01981 & 0.95626 & 0.88725 & 0.03517 & 0.91618 & 24.19711 & 12.35330 & 87.79542 & 5.00000 & 2.32456 & 10.97660 & 1.51910 & 0.48987 & 2.96390 \\
 & LL$_4$ & 0.93966 & 0.01808 & 0.95435 & 0.88619 & 0.03207 & 0.91268 & 23.53720 & 13.27045 & 114.15153 & 5.00000 & 1.95966 & 10.68782 & 1.57904 & 0.47235 & 2.92375 \\
\bottomrule
\end{tabular}}
\end{table*}

%\begin{figure}[htbp]
%  \centering
%  \includegraphics[width=0.4\textwidth]{images/dice_across_configurations.png}
%  \caption{Dice scores of CL, FL and average LL across each experimental configuration. LL is averaged over all clients.}
%  \label{fig:dice_across_configs}
%\end{figure}

\subsection{Visual Comparison of Results}
\begin{figure*}[htbp]
  \centering
  \includegraphics[width=\textwidth]{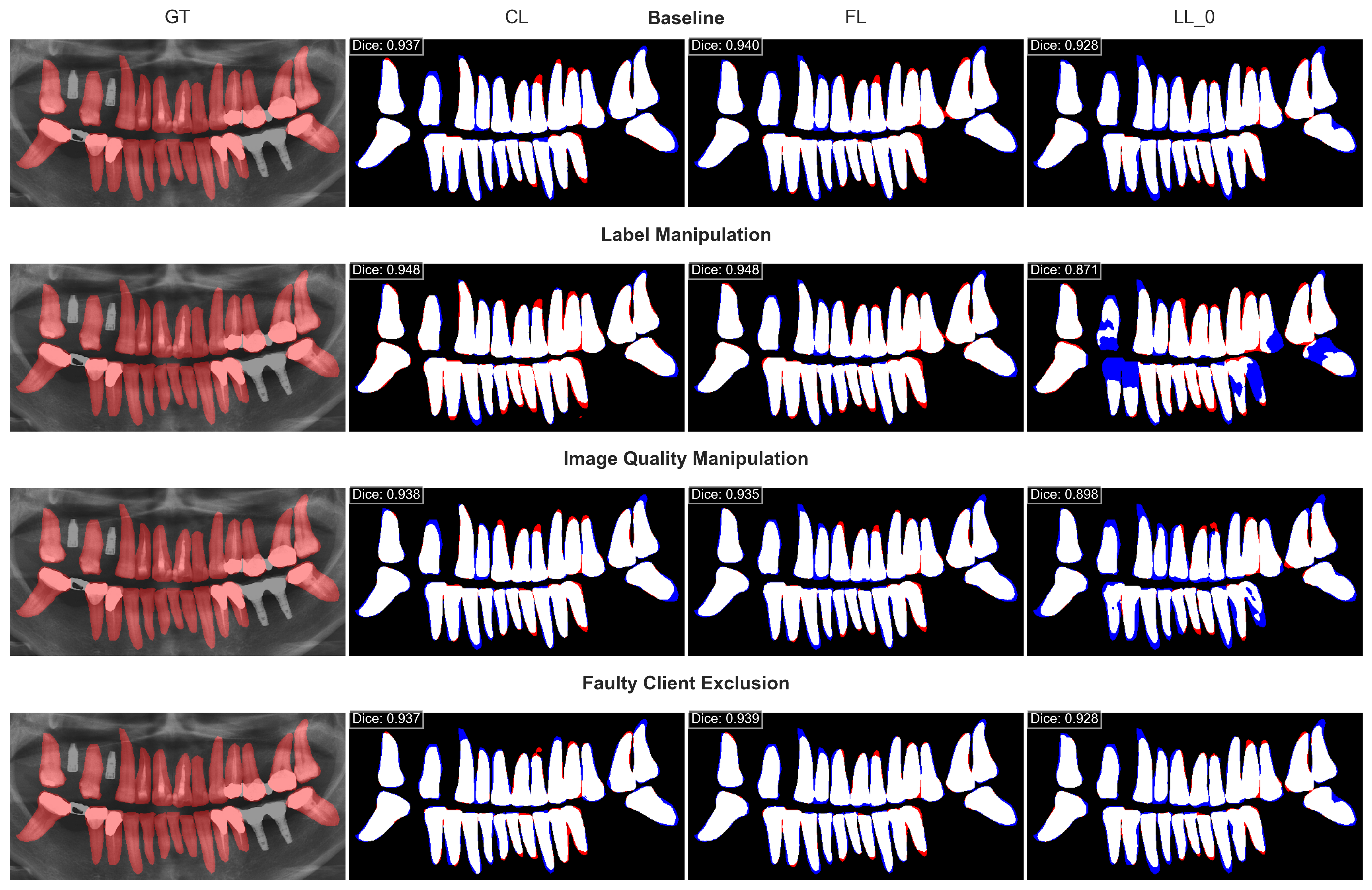}
  \caption{Prediction overlays for different experimental configurations and learning paradigms. Predictions were made on the full images, with Dice scores computed before cropping. For visualization, predictions and ground truths are cropped tightly around the labeled region. Difference masks use color coding: white = true positive, red = false positive, blue = false negative, black = true negative. Only one LL client is shown, representing the 'faulty client' in scenarios involving manipulations.}
  \label{fig:example_predictiongs_all_configs}
\end{figure*}

\begin{figure*}[htbp]
  \centering
  \includegraphics[width=0.9\textwidth]{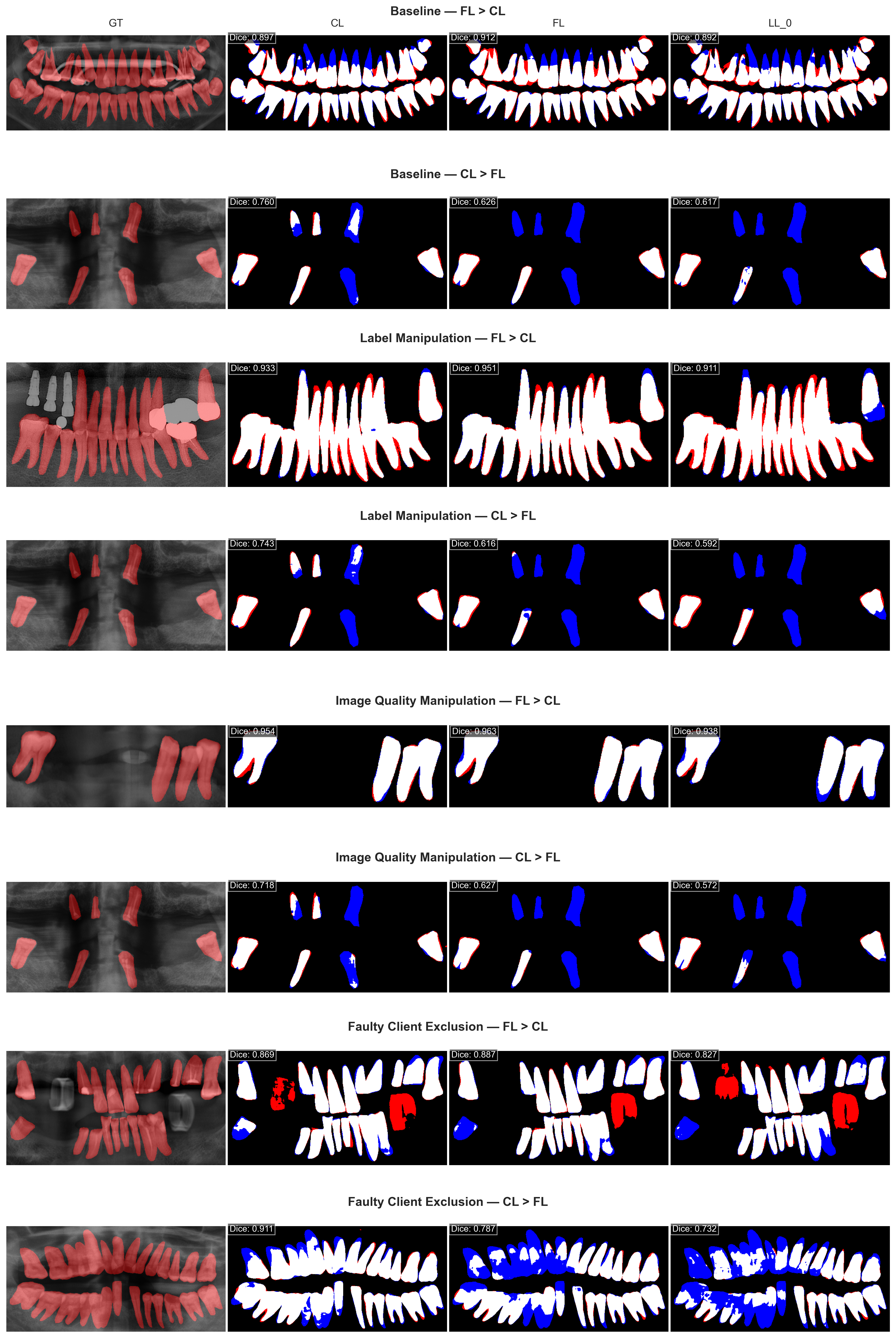}
  \caption{
        Visual comparison of the most extreme segmentation cases for each configuration. 
        For each configuration, the two most pronounced examples are shown: one where 
        FL outperforms CL ($FL > CL$) 
        and one where CL outperforms FL ($CL > FL$). 
        Each row shows the ground truth (GT) with overlayed mask, followed by the difference maps for 
        CL, FL, and the LL$_0$, with corresponding Dice scores. 
        True positives are shown in white, true negatives in black, false positives in red, and false negatives in blue.
    }
  \label{fig:example_predictiongs_extreme_cases}
\end{figure*}

Representative examples for all configurations and learning paradigms are shown in Figure \ref{fig:example_predictiongs_all_configs}, while Figure \ref{fig:example_predictiongs_extreme_cases} highlights the most extreme discrepancies between FL and CL. Overall, the visual comparisons indicate that CL and FL achieve highly similar segmentation performance across configurations, with small deviations found mostly in the apical regions. LL$_0$ consistently performs worse, mainly showing severe undersegmentation for some teeth.

The extreme case examples reveal that the largest performance gaps between CL and FL often arise in challenging cases, such as patients with sparse dentition, implants or crowns, or radiographs with artefacts. It also shows that FL errors are often spatially clustered, resulting in large contiguous false negative or false positive regions, whereas CL errors tend to be more scattered across the image. This pattern aligns with the impression from the Dice score distributions in Figure \ref{fig:all_boxplot}, where FL shows a slightly broader spread of scores than CL, suggesting greater sensitivity to outlier cases, despite achieving comparable or higher median performance.

\section{Discussion}
A systematic evaluation of each experimental configuration and its corresponding learning paradigm yielded several notable findings. Across all training processes, no significant overfitting was detected, and the hyperparameter selection was validated as appropriate for the dataset. In Figure \ref{fig:all_progress}, an immediate increase in training loss following round termination was observed for FL. This phenomenon is likely attributable to the model aggregation process; however, the stability of the validation loss suggests that the models retained robust generalization capabilities despite the transient training loss escalation.

In terms of overall performance for the baseline models, Figure \ref{fig:all_boxplot} demonstrates that the FL model achieved the significant highest performance, with the CL model following, and LL models exhibiting the poorest performance.

When manipulations to the labels were introduced, as illustrated in Figure \ref{fig:all_progress}, the faulty client (Client-0) consistently underperformed relative to the other clients. This pattern was evident in both the training loss trajectories across FL and LL paradigms. Furthermore, as shown in Figure \ref{fig:all_boxplot}, FL maintained its superior performance, followed by CL, whereas LL experienced a noticeable and significant decline in performance.

A comparable trend was observed under the image quality manipulation conditions. Figure \ref{fig:all_progress} indicates that Client-0 again underperformed compared to its peers in both FL and LL settings. For this experimental configuration the FL model significantly outperformed the other learning paradigms, with CL in second place and LL in third. \vspace{\baselineskip}

We found that clients whose data had been tampered with, either via label errors or added image noise, consistently exhibited elevated training losses and slower convergence compared to their peers, highlighted in red in Figure \ref{fig:all_progress}. This divergence in per-client training loss trajectories can therefore act as an anomaly detector: by monitoring each client’s loss curve against the cohort distribution (for example, flagging any client whose loss exceeds the median by more than a predefined threshold), one can automatically identify and temporarily exclude “faulty” participants from the federated aggregation. Interestingly, this elevated‐loss behavior does not manifest in the client’s validation‐loss curves, which converge similarly to those of the other participants. Moreover, when we re-evaluate the same client on its clean validation set (i.e. without label- or image quality manipulations), the validation loss decreases noticeably (see Figure \ref{fig:all_progress}). In practice, these “loss-curve fingerprints” could serve as safeguards against corrupted updates. This way, the faulty client can be excluded, and a small gain in overall performance can be achieved, with FL under the Faulty Client Exclusion configuration performing slightly better than FL with label manipulation from the faulty client. While these observations are promising, the underlying cause warrants further investigation. One plausible explanation involves the use of Batch Normalization within our U-Net blocks: during training, BatchNorm layers continually update their running mean and variance to accommodate the noisy samples, leading to a higher convergence plateau in the training loss. However, during validation, in which the BatchNorm statistics are fixed, the model’s parameters remain unchanged, yielding a validation, loss trajectory that mimics that of the uncorrupted clients. Finally, when evaluated on completely clean labels, the model’s learned feature filters suppress the impact of the previously encountered noise, resulting in a lower validation loss. Overall, these “fingerprints” appear to be a combined consequence of the data corruption pattern, model architecture, and training regimen. They open the possibility of designing robust exclusion criteria to detect and mitigate under-performing clients in federated learning systems.\vspace{\baselineskip}

Most noteworthy is that FL models trained on corrupted data significantly outperformed the baseline LL models, approximating the baseline CL models. The baseline FL model emerged as the best across all learning paradigms for all configurations. This finding suggests that the advantages inherent in a federated setup, such as improved generalization through collaborative training, can compensate for the inclusion of models trained on lower-quality data from some clients. At the same time, the increase in performance across all learning paradigms in the image-manipulated configuration suggests that the added Gaussian noise has some sort of generalizing effect, which has a confounding impact on the conclusions to be made, and relates to a completely different topic in regards of model generalizability due to data augmentation. Nevertheless, the usefulness of a federated setup is substantial and should be taken into account along with other advantages such as the distributed computational burden across multiple clients, which facilitates parallelized model training, thereby reducing overall training time. However, this approach necessitates that all clients possess adequate computational resources, specifically in terms of GPU capacity, which may impose economic constraints. Due to GPU memory limitations, a batch size of 4 was employed, although a larger batch size would generally be preferable.\vspace{\baselineskip}

Several recent studies have demonstrated the potential of federated learning in dental image analysis. Schneider et al. \cite{SCHNEIDER} applied FL, LL, and CL to a dataset of 4177 panoramic radiographs from nine centers, showing that FL significantly outperformed LL and approached CL performance on tooth segmentation tasks. Rischke et al. \cite{challenges3} further reviewed the opportunities and challenges of FL in dentistry, highlighting privacy preservation and cross‐site heterogeneity as key considerations. Our results in tooth segmentation under label‐ and image‐quality manipulation confirm these earlier findings: FL models not only surpass LL in all scenarios but also maintain robustness comparable to CL, even when some clients submit corrupted updates.

In the wider medical‐imaging field, several studies have tackled annotation noise and client heterogeneity within FL. Wu et al. \cite{FedA3I} introduced FedA3I, which weights client updates by estimated annotation quality to mitigate cross‐client noise, improving segmentation performance on heterogeneous datasets. Xiang et al. \cite{FedIA}'s FedIA explicitly models incomplete or noisy labels across clients, as a completeness-aware aggregation weight, based on predictions from a first-stage model and its loss progression. This allows it to adjust the aggregation accordingly to the quality of data at each client. Wicaksana et al. \cite{FedMix}’s FedMix goes further by supporting mixed supervision levels (pixel‐, box‐, and image‐level) in a unified FL framework, thereby enabling using different precision-level of labels for the same task, pixel-wise segmentation. Our observation, that corrupted clients can be automatically identified by their divergent training‐loss trajectories and safely excluded, resonates with these quality‐aware aggregation strategies and suggests a practical mechanism for enhancing FL robustness in dental and other medical imaging contexts.

The study was conducted using the Flower AI federated learning framework, which was used to simulate a distributed training environment with configurable parameters such as the number of clients, epochs training rounds, and client sampling strategies. For the present study, the standard IidPartitioner was employed in conjunction with FedAvg; however, future investigations could examine the impact of dataset heterogeneity, particularly with respect to status labels of teeth, on FL performance \cite{SCHNEIDER}, as well as the influence of alternative aggregation strategies (e.g., FedAdaGrad \cite{FedAdaGrad}, FedProx \cite{FedProx}, FedDM \cite{FedDM}) in combination with annotation completeness estimations. But also investigate the possibility of integration with enterprise-ready platforms such as NVFLARE to facilitate clinical implementation \cite{flower_nvflare}. 

\section{Conclusion}
In summary, our experiments confirm that federated learning not only matches centralized learning in accuracy but also exhibits greater resilience to labeling and imaging artifacts compared to centralized learning and local learning. By monitoring per-client loss to exclude faulty participants or adjust the weighting of these, FL can further improve segmentation performance. These findings substantiate FL’s potential as a practical, privacy-preserving approach for collaborative model training within the medical field.

\section{Competing interests}
No competing interest is declared.

\section{Author contributions statement}
%\begin{comment}
J.R., L.E., A.I. and R.P. conceived the experiment(s),  J.R. conducted the experiment(s), J.R. and R.P. analysed the results. J.R., K.N., S.J., L.E., A.I. and R.P. wrote and reviewed the manuscript.
%\end{comment}

\section{Acknowledgments}
%\begin{comment}
The authors thank the anonymous reviewers for their valuable suggestions. This work was funded by: (1) the Independent Research Fund Denmark, project “Synthetic Dental Radiography using Generative Artificial Intelligence”, grant ID 10.46540/3165-00237B, for computational resources; (2) the Aarhus University Centre for Digitalisation, Big Data and Data Analytics and the Circle U Seed Funding Scheme for salary costs.
%\end{comment}

\clearpage
\newpage
\begin{appendices}
\section{Appendix}\label{sec11}
\subsection{Appendix table - Statistical Significance Tests}\label{subsec3}
%% LaTeX Table: Within-Configuration
\begin{table}[htbp]
\caption{Within-Configuration Wilcoxon p‑values and significance (yes/no). Bonferroni‐adjusted $\alpha = 0.00278$ (over 18 valid tests).}
\label{tab:within-configuration}
\resizebox{\textwidth}{!}{%
\begin{tabular}{llrrrrrrrrrr}
\toprule
Configuration & Comparison & DICE p & DICE sig & IOU p & IOU sig & HD p & HD sig & HD95 p & HD95 sig & ASSD p & ASSD sig \\
\midrule
Baseline & LL$_0$ vs CL & 4.28e-34 & yes & 4.04e-34 & yes & 4.76e-07 & yes & 1.68e-28 & yes & 6.23e-31 & yes \\
Baseline & LL$_0$ vs FL & 1.02e-34 & yes & 9.05e-35 & yes & 2.74e-06 & yes & 2.13e-22 & yes & 4.79e-27 & yes \\
Baseline & CL vs FL & 3.43e-06 & yes & 2.91e-06 & yes & 8.82e-01 & no & 1.74e-01 & no & 9.52e-02 & no \\
Label Manipulation & LL$_0$ vs CL & 3.12e-34 & yes & 2.74e-34 & yes & 3.62e-05 & yes & 1.52e-29 & yes & 7.76e-32 & yes \\
Label Manipulation & LL$_0$ vs FL & 3.55e-34 & yes & 3.17e-34 & yes & 5.94e-14 & yes & 4.69e-33 & yes & 3.70e-33 & yes \\
Label Manipulation & CL vs FL & 3.81e-26 & yes & 1.96e-26 & yes & 3.38e-04 & yes & 1.04e-24 & yes & 4.68e-31 & yes \\
Image Quality Manipulation & LL$_0$ vs CL & 1.49e-35 & yes & 1.49e-35 & yes & 3.01e-16 & yes & 1.42e-34 & yes & 7.83e-35 & yes \\
Image Quality Manipulation & LL$_0$ vs FL & 1.49e-35 & yes & 1.49e-35 & yes & 4.48e-19 & yes & 1.66e-35 & yes & 1.49e-35 & yes \\
Image Quality Manipulation & CL vs FL & 1.90e-03 & yes & 1.85e-03 & yes & 2.56e-01 & no & 2.80e-01 & no & 3.60e-04 & yes \\
Faulty Client Exclusion & CL vs FL & 1.29e-10 & yes & 9.49e-11 & yes & 2.00e-01 & no & 2.43e-03 & yes & 4.00e-09 & yes \\
\bottomrule
\end{tabular}}
\end{table}

%% LaTeX Table: Within-Paradigm
\begin{table}[htbp]
\caption{Within-Paradigm Wilcoxon p‑values and significance (yes/no). Bonferroni‐adjusted $\alpha = 0.00278$ (over 18 valid tests).}
\label{tab:within-paradigm}
\resizebox{\textwidth}{!}{%
\begin{tabular}{llrrrrrrrrrr}
\toprule
Paradigm & Comparison & DICE p & DICE sig & IOU p & IOU sig & HD p & HD sig & HD95 p & HD95 sig & ASSD p & ASSD sig \\
\midrule
CL & Baseline vs Label Manipulation & 9.85e-25 & yes & 8.63e-25 & yes & 3.79e-07 & yes & 4.48e-31 & yes & 7.94e-35 & yes \\
CL & Baseline vs Image Quality Manipulation & 7.34e-03 & no & 7.50e-03 & no & 2.91e-01 & no & 7.65e-02 & no & 6.22e-02 & no \\
CL & Baseline vs Faulty Client Exclusion & 3.10e-02 & no & 3.23e-02 & no & 7.63e-01 & no & 4.81e-05 & yes & 1.27e-03 & yes \\
FL & Baseline vs Label Manipulation & 3.20e-04 & yes & 3.47e-04 & yes & 4.69e-02 & no & 6.57e-15 & yes & 2.76e-18 & yes \\
FL & Baseline vs Image Quality Manipulation & 2.14e-01 & no & 1.90e-01 & no & 1.56e-01 & no & 4.14e-06 & yes & 4.06e-06 & yes \\
FL & Baseline vs Faulty Client Exclusion & 2.23e-01 & no & 2.15e-01 & no & 8.78e-01 & no & 4.78e-01 & no & 1.10e-01 & no \\
LL$_0$ & Baseline vs Label Manipulation & 1.09e-24 & yes & 1.03e-24 & yes & 1.52e-04 & yes & 1.10e-29 & yes & 9.88e-31 & yes \\
LL$_0$ & Baseline vs Image Quality Manipulation & 4.66e-28 & yes & 2.67e-28 & yes & 8.93e-05 & yes & 7.47e-17 & yes & 9.69e-24 & yes \\
\bottomrule
\end{tabular}}
\end{table}
\end{appendices}

\clearpage
\newpage

\bibliography{references}

%USE THE BELOW OPTIONS IN CASE YOU NEED AUTHOR YEAR FORMAT.
%\bibliographystyle{abbrvnat}
%\bibliography{reference}

%% sample for biography with author's image
%\begin{comment}
\begin{biography}{%
  \includegraphics[width=55pt]{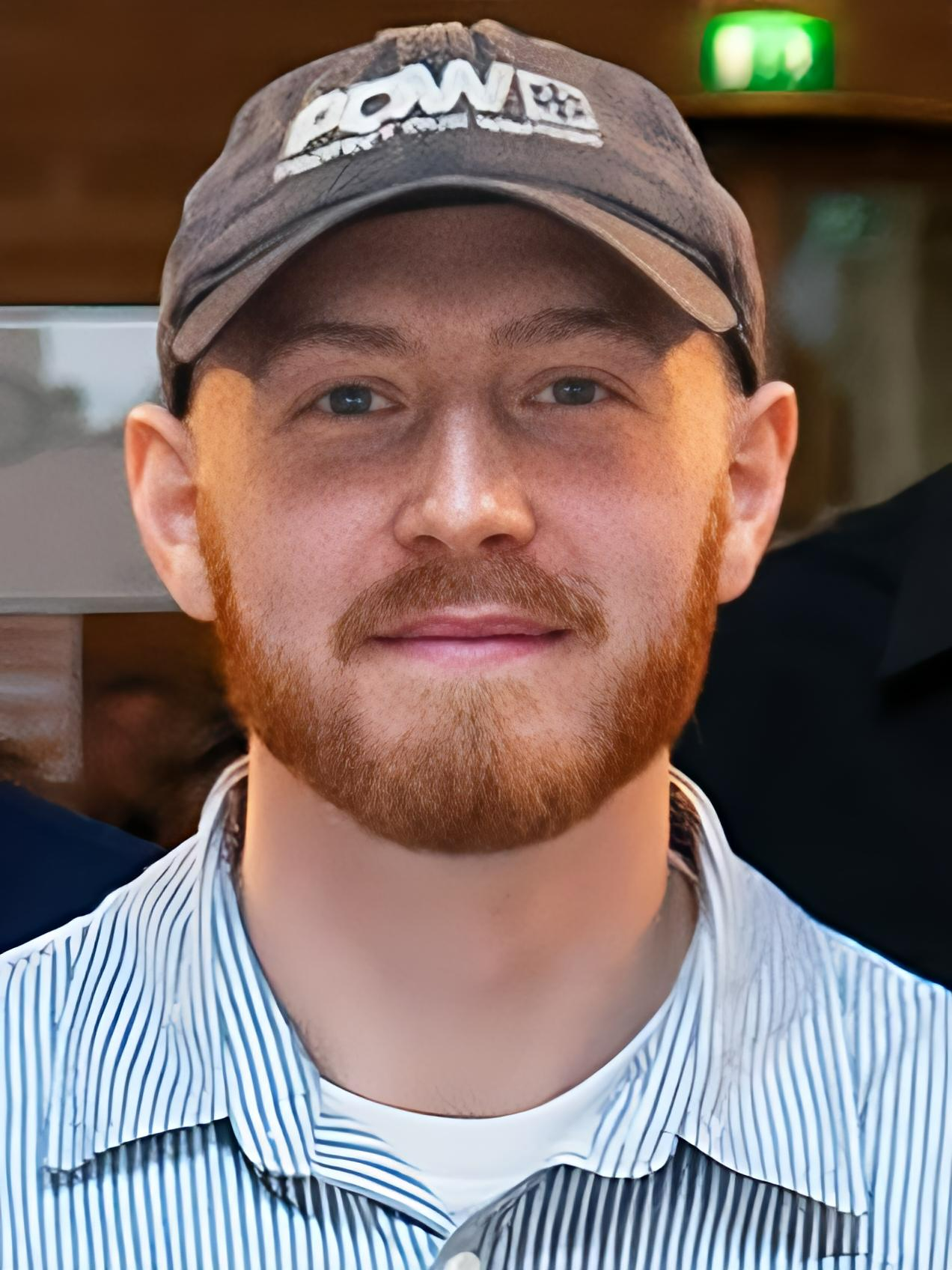}%
}{%
  \author{Johan Andreas Balle Rubak.} Biomedical Engineering student at Aarhus University and Research Assistant in the Department of Dentistry and Oral Health. I hold a deep interest in medical image modalities, 3D-modelling tools for clinical workflows and how to enable AI within these field, in particularly with federated learning.
}
\end{biography}
%\end{comment}
\end{document}